# Design of the Reverse Logistics System for Medical Waste Recycling Part II: Route Optimization with Case Study under COVID-19 Pandemic

Chaozhong Xue[#], Yongqi Dong[#], *Student Member, IEEE*, Jiaqi Liu, Yijun Liao, and Lingbo Li

*Abstract*— Medical waste recycling and treatment has gradually drawn concerns from the whole society, as the amount of medical waste generated is increasing dramatically, especially during the pandemic of COVID-19. To tackle the emerging challenges, this study designs a reverse logistics system architecture with three modules, i.e., medical waste classification & monitoring module, temporary storage & disposal site (disposal site for short) selection module, as well as route optimization module. This overall solution design won the Grand Prize of the "YUNFENG CUP" China National Contest on Green Supply and Reverse Logistics Design ranking 1st. This paper focuses on the design of the route optimization module. In this module, a route optimization problem is designed considering transportation costs and multiple risk costs (e.g., environment risk, population risk, property risk, and other accident-related risks). The Analytic Hierarchy Process is employed to determine the weights for each risk element, and a customized genetic algorithm is developed to solve the route optimization problem. A case study under the COVID-19 pandemic is further provided to verify the proposed model. Limited by length, detailed descriptions of the whole system and the other modules can be found at https://shorturl.at/cdY59.

## I. INTRODUCTION

In recent decades, with the expansion of medical waste, the mismatch between the amount of medical waste generated and the amount of disposal is gradually aggravated, especially during the outbreak of the Coronavirus Disease 2019 (COVID-19) pandemic. If medical waste is not properly managed and recycled, it might cause serious risks [1]. Thus, the public is gradually increasing their concerns about the management of medical waste on a global basis [2]–[7]. It is crucial to design a safe, reliable, and effective reverse logistics system for medical waste disposal and recycling.

Regarding the development of reverse logistics system for medical waste, most studies focus on network design [3], [4], [6], [8]–[12]. There are seldom studies that design an integrated reverse logistics system integrating second-level temporary storage & disposal site selection and reverse transportation route optimization. This study aims to fill such research gap with an advanced system architectural design and tackles the optimization of reverse logistics transportation in two phases, i.e., the temporary storage & disposal site (disposal site for short) selection phase, which is illustrated in another paper [13], and the route optimization phase, which is depicted in detail in this paper.

In terms of route optimization for medical waste recycling, Gao et al. [14] combined the vehicle routing problem of medical facilities and the collection problem of clinics' medical waste to the affiliated hospital and proposed a compact mixed-integer linear programming model considering the differentiated collection strategy and time windows to solve the problem. Their proposed method could obtain solution efficiency in solving large-scale problem instances. Shi et al. [15] paper developed a mixed integer linear programming model with minimizing costs for medical waste reverse logistics networks and validated the efficiency and practicability of the proposed model through an example dealing with medical waste returned from some hospitals to a given manufacturer. Gao and Ye [16] designed a multivehicle mixed delivery mode considering the fixed costs, transportation costs, carbon emission costs, and time penalty costs. Their experimental results show that the multivehicle mixed delivery model can reduce costs more effectively than single-vehicle models. Liu et al. [17] established an efficiency-of-transport model of medical waste between hospitals and temporary storage stations using an ant colony–tabu hybrid algorithm. They chose Wuhan City in China as an example to test their proposed model and achieved good verification. Xu et al. [18] proposed a multi-cycle and multi-echelon location-routing model integrating reverse logistics incorporating a negative utility objective function generated based on panel data and developed an improved algorithm named particle swarm optimization-multi-objective immune genetic algorithm to solve the model. This study provided firms with more flexible location-routing options by dividing them into multiple cycles and provided valuable management recommendations for logistics planning.

It is identified that available studies seldom consider the various risk costs aroused by medical waste, especially for the risks brought by waste with high infectivity. This study attempts to design a route optimization model considering transportation costs and multiple risk costs (e.g., environment risk, population risk, property risk, and other accident-related

[#] These authors contributed equally to this work and should be considered as co-first authors.

Chaozhong Xue is with Department of Traffic Engineering and Key Laboratory of Road and Traffic Engineering, Tongji University, Shanghai, 200092, China. (e-mail: 1750700@tongji.edu.cn).

Yongqi Dong is with Delft University of Technology, Delft, 2628 CN, the Netherlands, he is also with University of California, Berkeley, Berkeley, CA 94720, USA. (e-mail: yongqidong@berkeley.edu).

Jiaqi Liu is with Department of Traffic Engineering and Key Laboratory of Road and Traffic Engineering, Tongji University, Shanghai, 200092, China. (corresponding author, phone: +86 18019113619; e-mail: liujiaqi13@tongji.edu.cn).

Yijun Liao is with Department of Traffic Engineering and Key Laboratory of Road and Traffic Engineering, Tongji University, Shanghai, 200092, China. (e-mail: 1851021@tongji.edu.cn).

Lingbo Li is with Department of Traffic Engineering and Key Laboratory of Road and Traffic Engineering, Tongji University, Shanghai, 200092, China. (e-mail: 2233430@tongji.edu.cn).

risks). The Analytic Hierarchy Process is employed to determine the weights for each risk element, and a customized genetic algorithm is developed to solve the proposed route optimization problem. A case study under the COVID-19 pandemic using real-world data from Dalian City in China is also provided to verify the proposed model.

## II. REVERSE LOGISTICS SYSTEM ARCHITECTURE DESIGN

To tackle the aforementioned challenges and fill the research gaps in designing the reverse logistics system, this study proposes and designs a system architecture with three modules, i.e., waste classification & monitoring module, temporary storage & disposal site selection module, as well as route optimization module. The system architecture and the relationships between each module are illustrated in Figure 1. Detailed descriptions for each module are provided in another paper [13] which serves as the supplementary material and can be found at https://shorturl.at/cdY59.

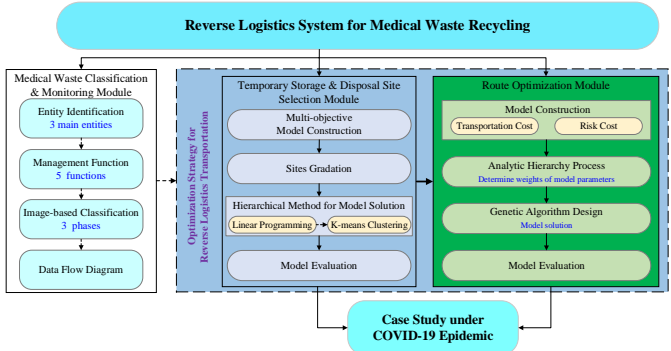

Figure 1. Illustration of the proposed reverse logistics system architecture

## III. ROUTE OPTIMIZATION FOR MEDICAL WASTE RECYCLING

This study models the route optimization problem for medical waste recycling integrating transportation costs and risk costs. To goal is to find vehicle routing paths covering all the temporary storage & disposal sites (determined by methods illustrated in [13]) while minimizing the total cost. Besides the common transportation cost, multiple types of risk costs (e.g., environment risk, population risk, property risk, and other accident-related risks) are considered. The optimization model is constructed with minimizing the weighted joint costs as the objective. The Analytic Hierarchy Process (AHP) method is adopted to determine the model parameter weights. A customized Genetic Algorithm (GA) is developed and employed to solve the model. Finally, a case study using real-world data during the COVID-19 pandemic is provided to verify the proposed model.

### A. Transportation Cost Modeling

*Fixed cost*: The fixed cost mainly refers to the cost incurred by the vehicle to complete the specified recycling task, including vehicle maintenance, repair cost, equipment depreciation, driver's salary, etc. Assuming the fixed cost for vehicle $h$ is $C_h$, then the total fixed cost, is calculated as:

$$C_F = \sum_{h=1}^{H} C_h Z_h \quad (1)$$

where $h \in \{1,2,...,H\}$, $H$ is the number of vehicles used for medical waste recycling transportation, $Z_h$ is the Boolean variable indicating whether the vehicle $h$ is being used.

*Vehicle transportation cost*: Vehicle transportation cost is the cost incurred by the vehicle during transportation. It includes load capacity related fuel costs and toll fees. The total vehicle transportation cost is expressed as follows:

$$C_{VT} = \sum_{i=1}^{N} \sum_{j=1}^{N} \sum_{h}^{H} C_{ij} D_{ij} Z_{ijh} \quad (2)$$

where $i$ and $j$ represents the temporary storage & disposal sites which is determined by methods illustrated in [13], and $i,j \in \{1,2,...,N\}$, $N$ is the number of disposal sites; $C_{ij}$ is the unit cost per distance from site $i$ to $j$, $D_{ij}$ stands for the distance between $i$ to $j$, and $Z_{ijh}$ is the Boolean variable indicating whether the vehicle $h$ is being used during the transportation process from $i$ to $j$.

*Penalty cost*: Due to the highly toxic, infectious, and hazardous nature of medical waste, the time and temperature requirements for recycling them are stringent. If the recycling transportation is overrun, it will not only result in more fuel consumption and emissions, together with reduced customer satisfaction, but also pose risks to the surrounding population, environment, and property. Therefore, the problem here is a typical Vehicle Routing Problem with Time Windows (VRPTW). Furthermore, considering the national laws and regulations that medical waste must be disposed of within 48 hours after generation, the third-party recycling company requires a strict time window to complete medical waste collection and transportation, thus this is a VRPTW with a hard time window, and the corresponding penalty cost, $CP$, is defined as:

$$CP(h, t_r^h) = \begin{cases} M(t_O - t_r), & t_r < t_O \\ 0, & t_O \le t_r \le t_R \\ M(t_r - t_R), & t_r > t_R \end{cases} \quad (3)$$

where $t_r^h$ is the time when the vehicle $h$ returns to the centralized disposal site; $[t_O, t_R]$ is the required time window with starting time $t_O$ and ending time $t_R$; $M$ is a large constant serving as a coefficient.

*Transportation cost model assumption*: To build the mathematical model, it is assumed that,

1) There is no capacity limit at the centralized disposal site, and all medical waste transported from the temporal disposal center can be recycled.

2) Transport vehicles are first departed from the centralized disposal site, then cruise the temporary disposal centers, and finally return to the centralized disposal site. The vehicles are of the same type and have the same carrying capacity.

3) The amount generated at each medical waste temporary storage and disposal center can be scientifically estimated based on historical data, and the amount generated at any center is less than the vehicle's load capacity. All recycling transport vehicles can serve more than one temporary storage & disposal center, and each center is only served by one vehicle.

4) Road traffic congestion levels typically vary with time, however, this paper assumes that the level of congestion is constant over a short period.

*Transportation cost model construction:* The minimization model construction for total transportation cost,

$TC$, considering the fixed cost, vehicle transportation cost, and the penalty cost with a hard time window is as follows:

$$\min TC = C_F + C_{VT} + CP =$$
$$\sum_{h=1}^{H} C_h Z_h + \sum_{i=1}^{N} \sum_{j=1}^{N} \sum_{h=1}^{H} C_{ij} D_{ij} Z_{ijh} + \sum_{h=1}^{H} CP(h, t_r^h) \quad (4)$$

S.T.

$$\forall\, i \in \{1,2,\dots,N\},\ \sum_{h=1}^{H}\sum_{j=1}^{N} Z_{ijh} \leq H \quad (5)$$

$$\forall\, h \in \{1,2,\dots,H\}, \forall\, i \in \{1,2,\dots,N\},\ \sum_{j=1}^{N} Z_{ijh} \leq 1 \quad (6)$$

$$\forall\, i \in \{1,2,\dots,N\},\ \sum_{h=1}^{H}\sum_{j=1}^{N} Z_{ijh} = 1 \quad (7)$$

$$\forall\, j \in \{1,2,\dots,N\},\ \sum_{h=1}^{H}\sum_{i}^{N} q_i Z_{ijh} \leq Q \quad (8)$$

$$T_O^h + \sum_{i=0}^{N}\sum_{j=0}^{N} Z_{ijh}(t_{ij} + t_{si}) \leq T_R^h,\ h \in \{1,2,\dots,H\} \quad (9)$$

$$\sum_{i=0}^{N}\sum_{j=0}^{N} Z_{ijh}(t_i + t_{ij}) = t_j,\ j \in \{1,2,\dots,N\} \quad (10)$$

$$C_{ij} \geq 0;\ q_i \geq 0;\ D_{ij} \geq 0 \quad (11)$$

$$Z_{ijh} \in \{0,1\}; \quad (12)$$

where $q_i$ is the amount of medical waste generated by temporal storage and disposal center $i$; $Q$ is the capacity of the vehicle; $t_i$ is the time that the vehicle arrives at $i$; $t_{ij}$ is the travel time from $i$ to $j$; $t_{si}$ is the stay time at $i$ (for loading task); $T_O^h$ is the time when the vehicle departs from the centralized disposal site; and $T_R^h$ is the time when the vehicle finishes the collection task and returns to the centralized disposal site.

### B. Risk Cost Modeling

***Environment risk***: If an accident occurs during medical waste recycling, it will inevitably cause pollution to the surrounding environment such as water, air, and soil, and influencing factors includes wind direction, topography, and hydrology. Wind direction and wind speed play a major role in air pollution, while topography, hydrology, geology, and precipitation play a major role in groundwater pollution, thus it is impossible to spread the pollution evenly in all directions. In this study, the environment risk, $RE_{ij}$, is calculated as the probability of medical waste recycling transportation accidents multiplied by the area or volume that may be affected by such accidents:

$$RE_{ij} = \sum_{i=1}^{n}\sum_{j=1}^{n} p_{ij} h_m \left( A_{ij}^1 \theta_{ij}^1 + A_{ij}^2 \theta_{ij}^2 + A_{ij}^3 \theta_{ij}^3 \right) \quad (13)$$

where $p_{ij}$ stands for the probability of medical waste transportation accidents; $A_{ij}^1$, $A_{ij}^2$, and $A_{ij}^3$ stand for, in the event of an accident happening at road segment $(i,j)$, the influenced areas regarding water bodies, atmosphere, and soil respectively; $\theta_{ij}^1$, $\theta_{ij}^2$, and $\theta_{ij}^3$ denote the Sherwood number, i.e., proportionality factor of the medical waste compared to the standard diffuse area, in the water body, atmosphere, and soil respectively; and $h_m$ indicates the harmful level of pollution caused by medical waste.

***Population risk***: Population risk is usually defined as the product of the transportation accident rate and the number of exposed populations in the area affected by the accident. The impact area is usually expressed in terms of the center of the accident point and the threatened radius. The number of exposed populations in the affected area is assessed by the average population density of the influenced region multiplied by the area of the region. Referring to the literature, the population risk, $RP_{ij}$, under time-varying conditions during medical waste recycling transportation is finally given as:

$$RP_{ij} = \sum_{i=1}^{n}\sum_{j=1}^{n} p_{ij} \delta_{ij} \pi \lambda^2 \beta_{ij} \quad (14)$$

where $\delta_{ij}$ denotes the fatal rate in the event of an accident and medical waste leaking; $\lambda$ indicates the radius of the area affected in the event of an accident; and $\beta_{ij}$ represents the average population density within the area of influenced road segment $(i,j)$.

***Property risk***: The property risk, $RA_{ij}$, during medical waste recycling transportation is calculated using the probability of medical waste transportation accidents multiplied by the property losses within the possible influenced area:

$$RA_{ij} = \sum_{i=1}^{n}\sum_{j=1}^{n} \frac{S_{ij}^m}{S_{ij}} \times p_{ij}(V_{ij}^1 + V_{ij}^2) \times \gamma_{ij} \quad (15)$$

where $V_{ij}^1$ and $V_{ij}^2$ represents the value of personal property and public property respectively, within the influenced road segment $(i,j)$ of the possible accident; $S_{ij}$ represents the area of road segment $(i,j)$; $S_{ij}^m$ stands for the area of the influenced road segment $(i,j)$; and $\gamma_{ij}$ indicates the severity level of property damage.

***Recycling transportation accident rate***: It is identified that all risks are related to the probability of medical waste recycling transportation accidents, i.e., recycling transportation accident rate, so it is vital to illustrate its calculation. The recycling transportation accident rate, $p_{ij}(t_i)$, is calculated as the product of the possible transportation accident rate per unit of time and the probability of leakage in the event of an accident, together with the travel time:

$$p_{ij}(t) = q_{ij}^t P(R|A)_{ij} \Delta t_{ij} \quad (16)$$

where $p_{ij}(t)$ denotes the possible incidence of medical waste leakage accidents on the road segment $(i,j)$ for a vehicle departing at moment $t$; $q_{ij}^t$ denotes the incidence of transportation accidents on road segment $(i,j)$ for a vehicle at time $t$; $P(R|A)_{ij}$ denotes the rate of medical waste leakage under an accident that has already occurred on road segment $(i,j)$; and $\Delta t_{ij}$ denotes the travel time the vehicle spends on the road segment $(i,j)$,

$$\Delta t_{ij} = \frac{L_{ij}}{v_{ij}} \quad (17)$$

where $L_{ij}$ denotes the length of the road segment $(i,j)$ and $v_{ij}$ denotes the average speed of the vehicle traveling on the road segment $(i,j)$.

***Risk cost model assumption***: It is assumed that,

1) The occurrence rate of transportation accidents remains constant over time.

2) Assuming that the ratio of the diffusion area of negative effects in the atmosphere, water body, and soil to the standard diffusion area is consistent.

***Risk cost model construction:*** When calculating the total risk costs during the medical waste recycling transportation process, this study considers environment, population, and property risks, as well as other risks caused by potential accidents. The total risk cost, $RC$, can be expressed as

$$RC = w_1 RP_{ij} + w_2 RA_{ij} + w_3 RE_{ij} + w_4 p_{ij} \quad (18)$$

where $w_1$, $w_2$, $w_3$, and $w_4$ are weighted parameters, and $w_4 p_{ij}$ stands for all other risks caused by potential accidents.

As the environment risk, population risk, property risk, and other risks are represented by different dimensions, this paper introduces three coefficients, $K_e$, $K_p$, and $K_o$ to transform all the risks into losses of capital. In this way, the total risk cost is transformed into:

$$RC = w_1 K_p RP_{ij} + w_2 RA_{ij} + w_3 K_e RE_{ij} + w_4 K_o p_{ij} \quad (19)$$

***Weighted parameters determination by AHP:*** In this paper, the Analytic Hierarchy Process (AHP) is employed to quantify each influencing factor in risk cost and determine the value of the weighted parameters $w_1$, $w_2$, $w_3$, and $w_4$. The corresponding AHP diagram is illustrated in Figure 2.

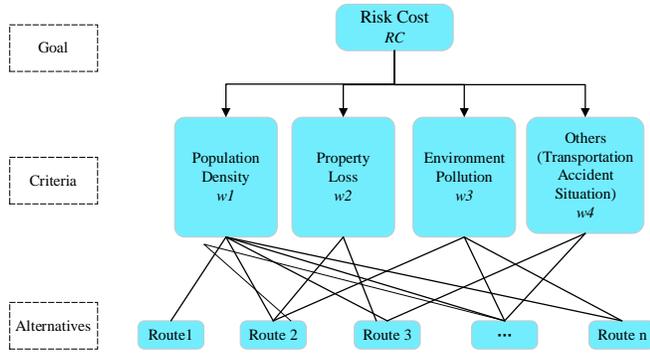

Figure 2. The diagram of the Analytic Hierarchy Process

TABLE I.    SCORING RESULTS FOR INFLUENTIAL FACTORS

| Influencing Factors | Population Density | Property Loss | Environment Pollution | Others |
|---|---|---|---|---|
| Population Density | 1 | 2 | 7 | 5 |
| Property Loss | 1/2 | 1/5 | 5 | 3 |
| Environment Pollution | 1/7 | 1/5 | 1 | 1/2 |
| Others | 1/5 | 1/3 | 2 | 1 |

Referring to the literature on medical waste and hazardous materials transportation, and taking into account the actual situation, the four affecting factors were scored using the "1-9 rating scale". The average scored results and illustrated in TABEL I. Pairwise comparison matrix (PCM) is obtained as

$$\text{PCM} = \begin{pmatrix} 1 & 2 & 7 & 5 \\ 1/2 & 1 & 5 & 3 \\ 1/7 & 1/5 & 1 & 1/2 \\ 1/5 & 1/3 & 1/2 & 1 \end{pmatrix} \quad (20)$$

Then the maximum eigenvalues and corresponding eigenvector (weights vector) of the PCM are obtained as

$\lambda_{\max} = 4.0201; W^T = (0.5267, 0.3005, 0.0630, 0.1098)$.

Carrying out the consistency test, the Consistency Index (CI) is calculated as

$$\text{CI} = \frac{\lambda_{\max} - n}{n-1} = \frac{0.0201}{3} = 0.0067 \quad (21)$$

where $n$ is the size of the comparison matrix and here $n = 4$.

As $\text{CI} < 0.1$, the PCM satisfies Saaty's consistency condition [19], which means the weights obtained by $W^T$, i.e., (0.5267, 0.3005, 0.0630, 0.1098) are acceptable.

*C. Integrated Route Optimization Model*

The final integrated route optimization model for medical waste recycling transportation is constructed by minimizing the joint costs of transportation and risk, obtained as:

$$\min Total\_Cost = (1-\alpha) * TC + \alpha * RC \quad (22)$$

where $\alpha$ is the proportional coefficient, which indicates the proportion of risk cost in the total cost. $\alpha$ is generally determined by problem requirements or expert experience.

*D. Model Variant under COVID-19 Pandemic Situation*

The COVID-19 epidemic brought new challenges to medical waste recycling transportation, e.g., high standards for waste storage, tight deadlines, and intensive tasks. Based on the regulations and new requirements, the integrated route optimization model illustrated by (22) needs to be customized and upgraded for the COVID-19 pandemic situation.

Vehicles used for transporting COVID-19-related medical waste are required to be disinfected using ultraviolet light, low temperature, or pharmaceutical sprays. This study assumes that the vehicles adopt the low-temperature sterilization method. Compared to ordinary vehicles, vehicles qualified for low-temperature sterilization require good sealing and insulation to ensure that medical waste is recycled and transported in a stable low-temperature condition. Thus, extra costs related to low-temperature sterilization are generated.

***Vehicle low-temperature sterilization cost***: The low-temperature sterilization cost consists of two parts, i.e., the cooling cost during vehicle travel and the cooling cost when loading and unloading.

The cooling costs of vehicle $h$ during travel, indicated by $C_{ZI}^h$, is calculated as

$$C_{ZI}^h = (1+\varepsilon) * R_c * S * \Delta T * (T_R^h - T_O^h) * U_C \quad (23)$$

where $\varepsilon$ indicates the degree of deterioration of the vehicle's carriage body; $R_c$ indicates the heat conductivity; $S$ indicates the average surface area of the carriage body; $\Delta T$ indicates the temperature difference between the inside and outside of the vehicle carriage body; and $U_C$ indicates the unit cooling cost.

The total cooling cost during all the vehicles' transportation travel, indicated by $TC_{ZI}$, is then calculated as

$$TC_{ZI} = \sum_{h=1}^{H} C_{ZI}^h \quad (24)$$

The cooling cost of vehicle $h$ when loading and unloading, indicated by $C_{ZS}^h$, is calculated as

$$C_{ZS}^h = (0.54 * Vol + 3.22) * \Delta T * U_C * \tau * \sum_{i=1}^{L} t_{si} \quad (25)$$

where $Vol$ indicates the volume of the carriage; $\tau$ is the factor indicating the average degree of the carriage door being opened, which is an empirical value. $\tau$ is estimated as 0.25 if no opening at all, 0.5 if opening once or twice per hour, 1.0 if opening three or four times per hour, 1.5 if opening five or six times per hour, and 2.0 if opening seven or eight times per hour. Thus, the total cooling cost for all the vehicles during their loading and unloading, indicated by $TC_{ZS}$, is then obtained as

$$TC_{ZS} = \sum_{h=1}^{H} C_{ZS}^{h} \qquad (26)$$

***Adjusted Penalty Cost***: The COVID-19-related medical waste generated during the pandemic is required to be disposed of within 36 hours (the holding time at hospitals cannot exceed 24 hours, while the staying time at medical waste disposal centers cannot exceed 12 hours). To meet the requirements and alleviate the risk when transporting COVID-19-related medical waste with high infectiousness, this study proposes to employ a "daytime + nighttime" mode, in which the general medical waste is cycled and transported during the daytime, while the COVID-19-related medical waste is only transported during the nighttime. Thus the penalty cost illustrated by (3) gets a strict time window with $t_O$ set as 18:00, and $t_R$ set as 24:00.

***Customized Model Variant under COVID-19 Pandemic Situation***: The customized total transportation cost under the COVID-19 pandemic situation by adding low-temperature sterilization cost, will be

$$TC_{COVID19} = C_F + C_{VT} + CP + TC_{ZI} + TC_{ZS} =$$
$$\sum_{h=1}^{H} C_h Z_h + \sum_{i=1}^{N} \sum_{j=1}^{N} \sum_{h=1}^{H} C_{ij} D_{ij} Z_{ijh} +$$
$$\sum_{h=1}^{H} CP(h, t_r^h) + \sum_{h=1}^{H} C_{ZI}^{h} + \sum_{h=1}^{H} C_{ZS}^{h} \qquad (27)$$

The risk cost will remain unchanged, thus the final customized route optimization model for medical waste recycling transportation under the COVID-19 pandemic situation is obtained by upgrading (23) into

$$\min\ Total\_Cost_{COVID19} = (1-\alpha) * TC_{COVID19} + \alpha * RC \quad (28)$$

*E. Model Solving Design Based on Genetic Algorithm*

A customized Genetic Algorithm (GA) is developed to solve the aforementioned model. The flow chart of the customized GA is illustrated in Figure 3. The GA algorithm cannot directly optimize the recycling transportation path, thus this study transforms the problem of optimizing the recycling transportation path into a problem regarding the combinatorial optimization of sequences. And natural number coding is employed to reduce the occurrence of invalid solutions.

***Customized Genetic Algorithm***

*1) Genetic encoding and population initialization*

Natural number encoding is employed as the genetic encoding method to transform the optimizing for recycling transportation path into the combinatorial optimization of sequences. Assume that the number of vehicles is $H$, the number of temporal storage & disposal center is $N$, the centralized disposal site is fixed at site $i = 29$ (and $i = 16$ for transporting COVID-19-related medical waste during nighttime); let $1, 2, 3, \cdots, N$ represents the temporal storage & disposal center, then the chromosome encoding with a length of $H + N + 1$ is generated to represent the vehicle routing path, with the sequence between two "29" as the sub-path for each vehicle.

For example, assuming there are $N$=9 temporal storage & disposal sites, and $H$=3 vehicles are used for transportation, then one chromosome coding as represented by the sequence

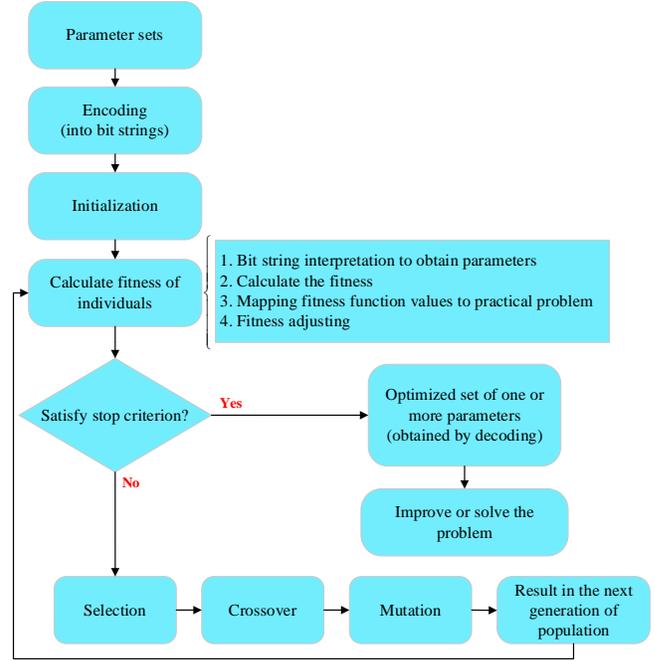

Figure 3. The flow chart of the customized Genetic Algorithm

"29-7-1-4-29-2-6-3-5-29-9-8-29" (length=3+9+1=14) can indicate three vehicles' paths, i.e., vehicle 1 departs from site 29 and visits sites 7, 1, and 4 sequentially and returns to 29, similarly, the path of vehicle 2 is "29 → 2 → 6 → 3 → 5 → 29"; and the path of vehicle 3 is "29 → 9 → 8 → 29".

With the proposed encoding method, the conditions illustrated in equations (5)-(12) can be easily checked. For instance, if the total amount of waste generated along the sites in a certain sub-path exceeds the capacity of the transportation vehicle, then it will violate the condition (8), and thus this specific chromosome will not be a successful candidate.

For population initialization, the initial population with a size of 30 is generated randomly.

*2) Fitness function*

The fitness function shows the strengths or weaknesses of a candidate solution. The value of the fitness for each chromosome determines the probability of its being kept in the next generation, and the larger the fitness function value is the higher the possibility will be. Since the goal of this study is to find the medical waste recycling transportation route paths minimizing the risk and cost, the fitness function can be set as the total integrated cost illustrated in (22) or ((28) under the COVID-19 pandemic situation):

$$fitness\_fun = Total\_Cost = (1-\alpha) * TC + \alpha * RC\ (29)$$

*3) Genetic operation*
   **a. Selecting the operator**

In this step, the roulette wheel operator is employed.

   **b. Crossover operator**

The crossover operation is for the rearrangement of homologous chromosomes into new populations and adding new characteristics to the original base. Partially mapped crossover (PMX) is employed. The positions of the crossover operators are randomly swapped and the genes perform the same operation in the same way to obtain new offspring individuals.

### c. Mutation Operator

Considering the characteristics of the recycling transportation optimization, the inversion mutation operator is employed in this paper.

### d. Termination rules

Regarding the nature of optimizing medical waste recycling transportation, to balance the solution performance and computational time, the termination rule is set as either the total cost improvement in the current iteration is less than 0.00001% compared with the total loss of the last iteration or a number of 300 iterations is reached. With this criterion, when the GA solution process is terminated, the paths encoded in the corresponding chromosome with the best fitness are selected as the optimal solution.

## IV. CASE STUDY

To verify the effectiveness of the proposed route optimization model, a case study is carried out using real-world empirical data collected from Dalian, a city in northern China. Medical units within two districts, Xigang District and Zhongshan District of Dalian City, were finally selected as the research target areas. There are a total of 112 medical units in the selected area, including 21 Primary (and above) hospitals and 91 small private hospitals and outpatient clinics. to crawl the specific information of the target Dalian city area. Baidu map open platform Application Programming Interface (API) is utilized to crawl the specific information containing detailed locations of these 112 medical units with latitude and longitude. As illustrated in [13], 28 temporary storage & disposal sites are selected, thus in this study, the objective is to find paths starting from the centralized disposal site (i.e., $i = 29$) cruising all the 28 sites, and return to the centralized disposal site while minimizing the total cost.

### A. Relevant Parameter and Parameter Assignments

In this study, the route planning function of Baidu Maps is employed to get the distance and travel time from one temporary storage & disposal site to another. Other parameters were determined by reviewing relevant literature, reports, and yearbooks, and reasonable assessments for the relevant model parameters are listed in TABLE II.

### B. Model Solution and Results Analysis

With the estimated model parameters listed in TABLE II, the medical waste recycling transportation route optimization model illustrated by (19) can be customized in terms of the real-world conditions of Dalian City. And the constructed model is then solved by the proposed GA algorithm described in section *III-E*.

Under the proposed "daytime + nighttime" transportation mode, two categories of the route paths are generated, i.e., paths for general medical waste recycling and transportation during the daytime, illustrated in Figure 4 (a), and paths for the COVID-19-related medical waste recycling and transportation

TABLE II. MODEL PARAMETERS ESTIMATION

| Parameters | Estimated Value |
| --- | --- |
| $C_h$ | 120 CNY |
| $C_{ij}$ | 20 CNY/ton/km |
| $Q$ | 3 ton |
| $\varepsilon$ | 0.08 |
| $S$ | 59.12m$^2$ |
| $\Delta T$ | 20 °C |
| $U_c$ | 0.5/h/m$^3$ |
| $Vol$ | 26.78m$^3$ |
| $R_c$ | 2.5 |
| $\tau$ | 2 |
| $K_e$ | 300 CNY/m$^3$ |
| $K_p$ | 10$^6$ CNY/person |

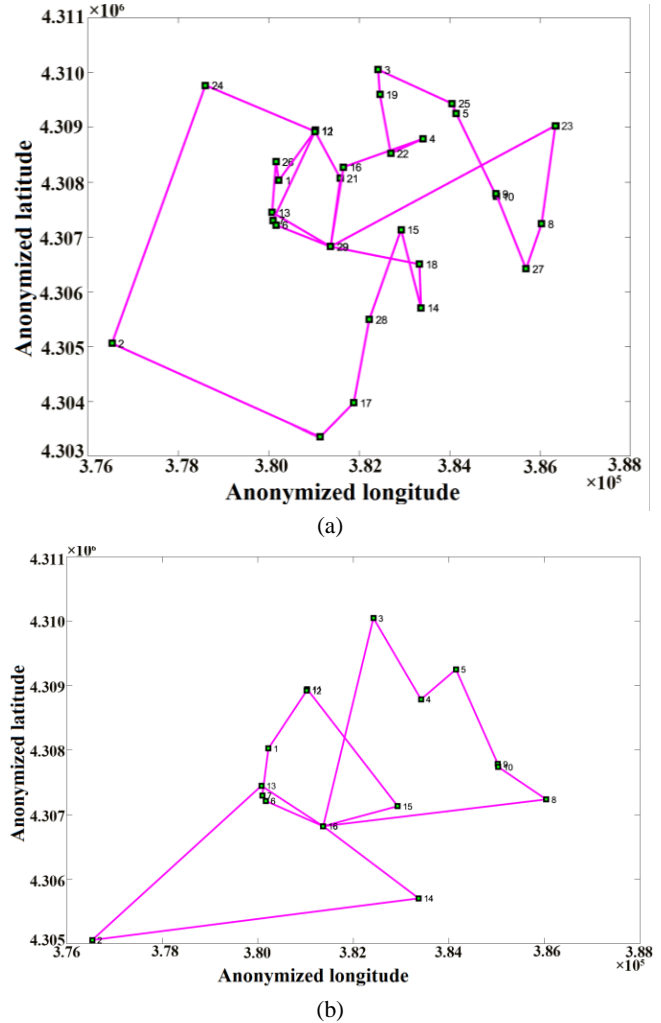

Figure 4. The illustration of optimized transportation paths: (a) paths for general medical waste transportation during the daytime, and (b) paths for COVID-19-related medical waste transportation during the nighttime

during the nighttime illustrated in Figure 4 (b). It is identified three vehicles are required for the general medical waste transportation during daytime with the three optimized paths "29-18-14-15-28-17-20-2-24-12-13-29", "29-23-8-27-10-9-5-25-3-19-22-4-16-29", and "29-21-11-1-26-7-6-29". The average transportation time is calculated as 3.17 hours per vehicle. While, for COVID-19-related medical waste transportation during the nighttime, also three vehicles are

required to serve three paths, i.e., "16-15-11-12-1-7-6-16", "16-3-4-5-9-10-8-16", and "16-14-2-13-16". Here, only the first 16 sites are involved in COVID-19-related medical waste recycling, and the centralized disposal site is site 16. The average transportation time is obtained as 3.24 hours per vehicle.

Since the actual routing paths during the epidemic were not available, in this paper, the randomly selected routes were selected as baselines, and the average results calculated from 10 iterations of randomly selected routes served as the benchmark for comparative evaluation.

For the daytime transportation of general medical waste, it is estimated the total converted cost of the benchmark routing paths is 6,371.42 CNY, and the transportation time is 10.6 hours, while with the proposed method, the total converted cost obtained by (28) is 3,431.42 CNY and the transportation time is 9.5 hours, which saves by 46.1% and 10.4% respectively. For the nighttime transportation of COVID-19-related medical waste, the benchmark of total converted cost is 4,330.07 CNY, and transportation time is 10.08 hours; with the proposed method, the total cost is reduced to 3,050.44 CNY, and the transportation time is reduced to 9.72 hours, which saves by 29.6% and 3.5% respectively. Furthermore, it is roughly estimated that the responding logistics company normally collected 9,503.60 tons of general medical waste per day within the specified hard time window, after employing the designed optimization model with the proposed "daytime + nighttime" mode, it can at least collect 7,662.40 more tons of COVID-19-related medical waste per day, delivering an increase of 80.6%. This can effectively solve the problem brought on by the surge of medical waste during the pandemic.

## V. CONCLUSION

Under the integrated reverse logistics system architecture with three modules, i.e., waste classification & monitoring module, temporary storage & disposal site selection module, as well as route optimization module, this paper elaborates on the modeling and solving of the route optimization problem for medical waste recycling transportation. Transportation costs and low-temperature sterilization costs under the COVID-19 pandemic situation, together with multiple types of risk costs are integrated to build the total cost for the route optimization problem, with Analytic Hierarchy Process (AHP) method adopted to estimate the risk-related model parameter weights. A customized genetic algorithm is designed and employed to solve the model. A case study using real-world data during the COVID-19 pandemic is provided to verify the proposed method. Compared with the benchmark, the proposed method can greatly reduce the total cost and transportation time, while dramatically increasing the total amount of medical waste the logistics system can transport.

Regarding future work, with the advancement of information technology, real-time data collection and monitoring of medical waste generation are becoming feasible, so it is recommended to consider the dynamic patterns of medical waste generation and integrate the prediction of medical waste generation at different sites into the modeling of route optimization. Thus, the route optimization model performance can be further advanced.